\documentclass[10pt,conference]{IEEEtran}
\IEEEoverridecommandlockouts
\usepackage{cite}
\usepackage{amsmath,amssymb,amsfonts}
\usepackage{algorithmic}
\usepackage{graphicx}
\usepackage{textcomp}
\usepackage{enumitem}
\usepackage{xcolor}
\def\BibTeX{{\rm B\kern-.05em{\sc i\kern-.025em b}\kern-.08em
    T\kern-.1667em\lower.7ex\hbox{E}\kern-.125emX}}

\usepackage{tcolorbox}
    
\begin{document}

\title{STaleX: A Spatiotemporal-Aware Adaptive Auto-scaling Framework for Microservices\\}

\author{\IEEEauthorblockN{Majid Dashtbani, Ladan Tahvildari}
\IEEEauthorblockA{\textit{Electrical and Computer Engineering} \\
\textit{University of Waterloo}\\
Waterloo, Canada \\
\{mdashtbani, ladan.tahvildari\}@uwaterloo.ca}
}

\maketitle

\begin{abstract}
While cloud environments and auto-scaling solutions have been widely applied to traditional monolithic applications, they face significant limitations when it comes to microservices-based architectures. Microservices introduce additional challenges due to their dynamic and spatiotemporal characteristics, which require more efficient and specialized auto-scaling strategies. Centralized auto-scaling for the entire microservice application is insufficient, as each service within a chain has distinct specifications and performance requirements. Therefore, each service requires its own dedicated auto-scaler to address its unique scaling needs effectively, while also considering the dependencies with other services in the chain and the overall application. This paper presents a combination of control theory, machine learning, and heuristics to address these challenges. We propose an adaptive auto-scaling framework, STaleX, for microservices that integrates spatiotemporal features, enabling real-time resource adjustments to minimize SLO violations. STaleX employs a set of weighted Proportional-Integral-Derivative (PID) controllers for each service, where weights are dynamically adjusted based on a supervisory unit that integrates spatiotemporal features. This supervisory unit continuously monitors and adjusts both the weights and the resources allocated to each service. Our framework accounts for spatial features, including service specifications and dependencies among services, as well as temporal variations in workload, ensuring that resource allocation is continuously optimized. Through experiments on a microservice-based demo application deployed on a Kubernetes cluster, we demonstrate the effectiveness of our framework in improving performance and reducing costs compared to traditional scaling methods like Kubernetes Horizontal Pod Autoscaler (HPA) with a 26.9\% reduction in resource usage.
\end{abstract}


\section{Introduction}
The rapid growth of internet users and the increasing demand for online services require flexible, scalable applications to handle fluctuating loads. Cloud computing provides a solution with scalable infrastructure and auto-scaling, which dynamically adjusts resources based on demand.

However, traditional monolithic architectures struggle to take advantage of cloud auto-scaling due to tightly coupled components. Microservices, by decomposing applications into independently deployable services, offer a more scalable solution\cite{b1,b2,b3}, but introduce challenges in scaling due to their spatiotemporal features \cite{b4,b5}.

These features include Service Level Objectives (SLOs), such as response times, which must be upheld across interconnected service chains \cite{b6}. A performance issue in one service can cascade, violating the SLOs of the entire chain \cite{b7}. Temporal features like workload fluctuations \cite{b8} and service invocation patterns, often overlooked by traditional scaling algorithms \cite{b9, b10, b11}, must be integrated into new strategies for improved effectiveness.

Spatial features such as service dependencies and specific resource needs complicate scaling. For example, service failures can disrupt dependent services, and variations in computational needs or initialization delays during scaling require service-specific scaling parameters \cite{b7}.

Addressing these challenges is critical for developing effective scaling strategies and enhancing the performance of microservice-based applications in cloud environments.

\textbf{Challenge.} Auto-scaling microservices presents several critical challenges in terms of performance, cost, and resilience:

\begin{itemize}[leftmargin=0.3cm]
    \item \textit{SLO Violations}: Interconnected services in a microservice architecture can cause performance degradation in one service to affect the entire service chain, making it difficult to maintain collective SLOs.
    
    \item \textit{Spatial Features}: Unique spatial characteristics, such as service dependencies and varying resource requirements, complicate the allocation of resources to meet SLOs effectively.
    
    \item \textit{Temporal Features}: Workload fluctuations and unpredictable traffic patterns further complicate auto-scaling, as sudden spikes can overwhelm services, leading to SLO violations.
    
    \item \textit{SLO and Cost Trade-Off}: Balancing SLOs with cost is difficult in microservices. Over-provisioning leads to higher costs, while under-provisioning risks performance degradation and SLO violations. Performance bottlenecks in one service can render resource allocations to others ineffective. 
\end{itemize}


\begin{figure*}[t]
\centering
\includegraphics[width=0.9\textwidth]{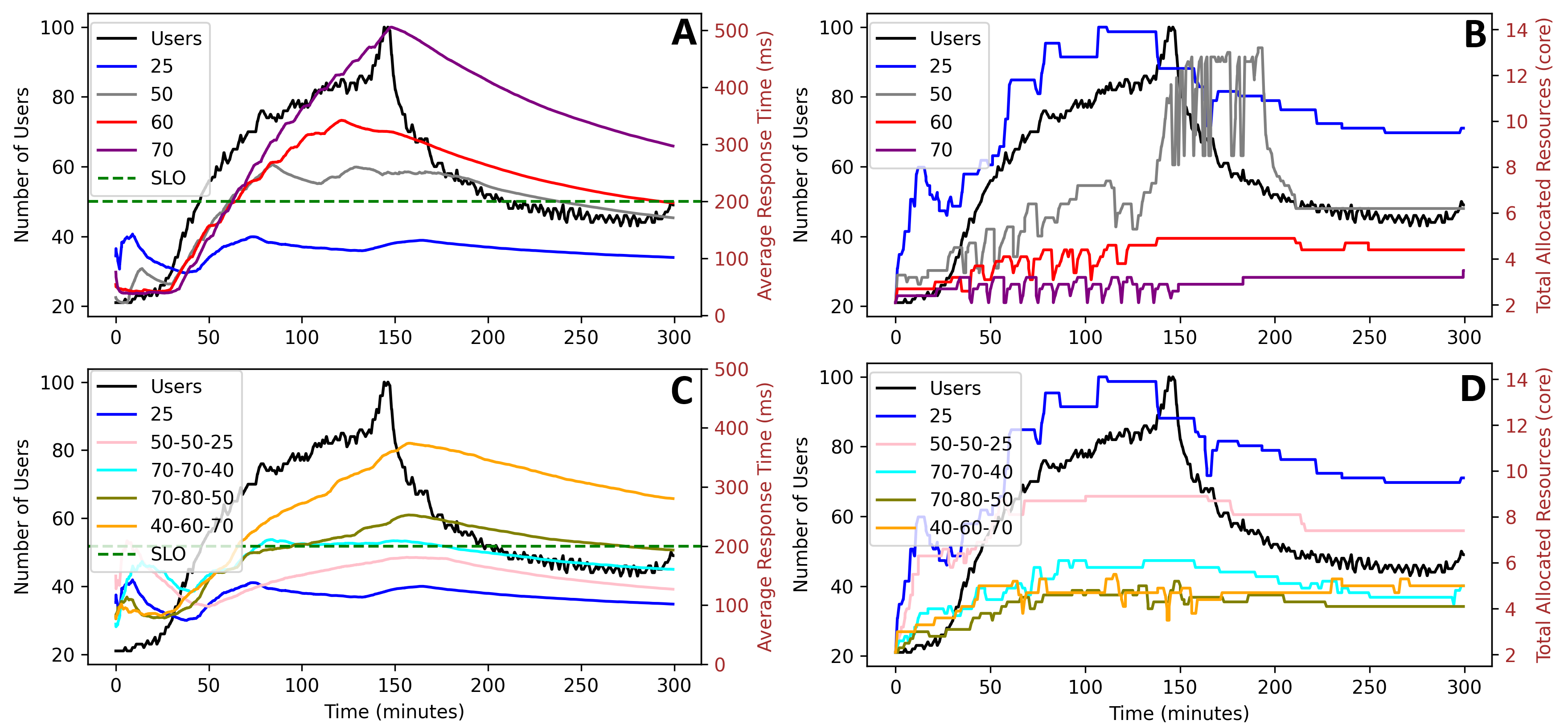}
\label{fig=HPAmicro}
\caption{A) Uniform Scaling - Response Time, B) Uniform Scaling - Resource Usage, C) Differentiated Scaling - Response Time, D) Differentiated Scaling - Resource Usage }
\end{figure*}

This study investigates the following research questions:

\textbf{RQ1: How do spatial features impact auto-scaling in microservices architectures?} This question examines the role of spatial features such as service dependencies, service role within the service chain, and service timing in improving the auto-scaling performance in microservices environments.

\textbf{RQ2: How do spatial and temporal features improve auto-scaling in microservices architectures?} This question investigates the combined effect of spatial features and temporal features, particularly workload prediction using Long Short-Term Memory (LSTM) models, on the efficiency and effectiveness of auto-scaling decisions in microservices applications.

The remainder of this paper is structured as follows: Section 2 presents a motivating example, Section 3 introduces our framework, Section 4 covers the evaluation, Section 5 discusses related work, and Section 6 concludes with future work.


\section{Motivating Example}
In this section, we demonstrate the potential of our framework through a motivating experiment. We deployed the Sock-Shop\cite{b12} demo microservice application on a self-hosted Kubernetes\cite{b13} cluster, using the Kubernetes Horizontal Pod Autoscaler (HPA)\cite{b14} to apply reactive auto-scaling. The goal is to evaluate HPA’s effectiveness in meeting the performance requirements of a SLO.

We focus on the '/login' service, which depends on a chain of microservices: 'front-end', 'user', and 'carts'. We define an SLO for '/login', requiring response times to stay under 200 ms. A limitation of HPA is that it primarily relies on low-level metrics, such as CPU and memory usage, and does not natively support high-level metrics like response time. To simulate realistic traffic, we used LOCUST.io\cite{b15} to generate a workload based on the WorldCup98\cite{b16} dataset.

In the first step, we applied the same CPU utilization threshold to all services to see if HPA can effectively scale while meeting the SLO. We tested various thresholds—25\%, 50\%, 60\%, and 70\%—to assess their impact on response time. Figure 1.A shows the response time at each threshold, while Figure 1.B illustrates resource usage over time.

In the second step, we set different thresholds for each service to evaluate whether tailored scaling improves performance and resource utilization. We tested a set of values for each service, focusing on dynamic scaling based on individual service needs. Figure 1.C shows the response time with varying thresholds for each service, while Figure 1.D illustrates resource usage distribution.

\begin{table}[htbp]
\caption{Comparison of Resource Usage and Response Time Violations for Uniform and Differentiated Scaling Approaches}
\begin{center}
\begin{tabular}{|c|c|c|c|c|c|}
\hline
\textbf{}&\multicolumn{3}{|c|}{\textbf{HPA Threshold (CPU Utilization \%)}}& \textbf{Resource} & \textbf{Violations} \\
\cline{2-4} 
\textbf{\#} & \textbf{Fron-end}& \textbf{User}& \textbf{Carts} & (Core/Min) & (ms)\\
\hline

1 & 25 & 25 & 25 & 4039 & 0  \\
\hline
2 & 50 & 50 & 50 & 2678 & 40225 \\
\hline
3 & 60 & 60 & 60 & 1610 & 77236 \\
\hline
4 & 70 & 70 & 70 & 1110 & 212760 \\
\hline
5 & 50 & 50 & 25 & 3630 & 3561 \\
\hline
6 & 70 & 70 & 40 & 2155 & 42561 \\
\hline
7 & 70 & 80 & 50 & 1886 & 50492 \\
\hline
8 & 40 & 60 & 70 & 1621 & 131549 \\
\hline

\end{tabular}
\label{HPA_Results}
\end{center}
\end{table}

\begin{figure*}[t]
\centering
\includegraphics[width=0.9\textwidth]{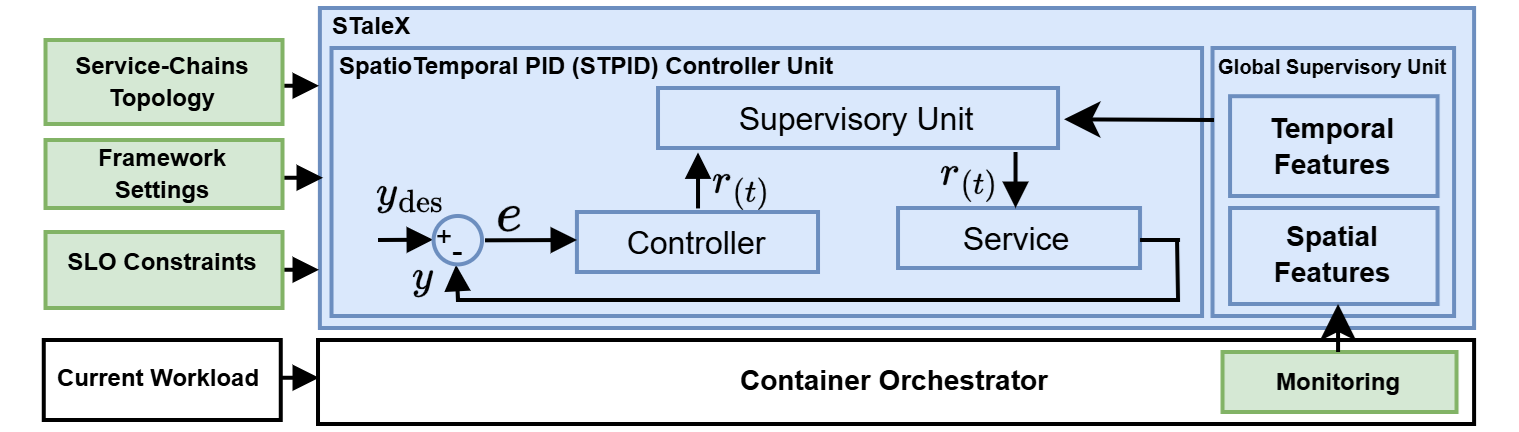}
\label{fig=framework}
\caption{STaleX framework: Green boxes represent the framework inputs, and blue boxes illustrate the framework components }
\end{figure*}

Table I summarizes the results, comparing CPU utilization thresholds, total resource usage (in cores/min), and response time violations (in ms). With uniform thresholds (rows 1–4), increasing the CPU utilization from 25\% to 70\% led to significant response time violations, with the 70\% setting causing a large violation over 200 ms. In contrast, using different thresholds (rows 5–8) resulted in a more balanced approach, with tailored scaling improving both resource efficiency and response time.

The experiments highlight several key challenges in auto-scaling microservices. First, relying solely on low-level metrics like CPU utilization does not guarantee high-level SLOs, such as maintaining response time under 200 ms, as sudden workload spikes can still lead to violations. Additionally, using uniform thresholds across services fails to address varying resource needs and service dependencies, often resulting in inefficiencies and cascading violations through the service chain. Finally, balancing the trade-off between meeting SLOs and controlling costs is challenging, emphasizing the need for more adaptive, dynamic scaling strategies.


\section{Proposed Method}
In this section, we present our proposed framework for adaptive auto-scaling in microservices architectures, leveraging PID controllers to address SLO violations while considering spatial and temporal features.

\subsection{Problem Formulation}\label{AA}
\textbf{Definition.}  
Let \( A \) represent a \textbf{microservice application} consisting of a set of \textbf{services} \( \{ S_1, S_2, \dots, S_n \} \). Each service \( S_i \) is characterized by two vectors: the \textbf{service chain vector} \( C_i = \{ m_1, m_2, \dots, m_k \} \), which represents the sequence of microservices \( m_1, m_2, \dots, m_k \) that make up the service chain for \( S_i \), and the \textbf{replica vector} \( R_i(t) = \{ r_1, r_2, \dots, r_k \} \), which represents the number of replicas allocated to each corresponding microservice in the chain at time \( t \). Each service \( S_i \) also has a corresponding user-defined \textbf{\( \text{SLO}(S_i) \)}, which defines the desired performance, such as response time or throughput, for the service.

Each microservice \( m_i \) in the chain has two vectors of features: the \textbf{spatial feature vector} \( \vec{f}_{spatial}(m_i) \) and the \textbf{temporal feature vector} \( \vec{f}_{temporal}(m_i) \). The spatial feature vector \( \vec{f}_{spatial}(m_i) \) includes characteristics such as the service's role in the service chain, its dependencies, and resource specifications (e.g., CPU, memory, etc.). The temporal feature vector \( \vec{f}_{temporal}(m_i) \) contains attributes related to workload dynamics, such as workload prediction, service boot time, and traffic patterns.

\textbf{Problem.}  
The objective is to minimize the total \textbf{SLO violations} across all services by dynamically adjusting the resource allocation of microservices in the service chain. The key challenge lies in efficiently tuning the resources based on the \textbf{spatiotemporal features} of the system. 

Given a set of services \( S = \{ S_1, S_2, \dots, S_n \} \) with corresponding \( \text{SLO}(S_i) \), and a chain of microservices \( C_i = \{ m_1, m_2, \dots, m_k \} \), the goal is to minimize the total SLO violation, \( \delta(S_i) \), where \( \delta(S_i) \) represents the deviation from the desired SLO of service \( S_i \). 
The objective is to minimize this total violation by dynamically adjusting the resource allocations \( {R}_i(t) \), considering the spatiotemporal features of the system. The problem is formally expressed as:

\begin{equation}
\min_{\{ \vec{R}_1, \vec{R}_2, \dots, \vec{R}_n \}} \sum_{i=1}^{n} \delta(S_i)
\end{equation}

subject to system constraints such as resource limits, service dependencies, and cost considerations. The goal is to find the optimal resource allocation \( {r}_i(t) \) at time \( t \) for each microservice that minimizes the total SLO violation while satisfying system constraints. All the numbers involved in the formulation are natural numbers.

\subsection{Framework Overview}
Our proposed framework, Figure 2, consists of two main components: a \textbf{global supervisory component} and a set of \textbf{SpatioTemporal PID (STPID) controller units}. The framework takes in three key inputs: \textbf{Service-Chains Topology}, which defines the list of services and their dependencies within each chain; \textbf{SLO Constraints}, specifying performance targets for each service; and \textbf{Framework Settings}, which include parameters for controlling resource thresholds and the timing of auto-scaling decisions. In the following, we explain the role of each component in the scaling process.

\subsection{Global Supervisory Component}
The Global Supervisory Component is responsible for providing the spatiotemporal auto-scaling information of the entire microservice application to the Spatiotemporal PID (STPID) controller units. The following subsections discuss how the component incorporates both spatial and temporal features to ensure efficient and adaptive scaling.


\subsubsection{Spatial Features}
The Global Supervisory Component maintains a list for each service chain, which includes a vector of service utilization over a defined duration (e.g., 5 minutes). This duration is an input to the framework, allowing the user to define the time window for monitoring resource utilization. By tracking resource usage within each chain, the component can evaluate the state of each service in relation to others. This spatial awareness enables the system to adjust resources effectively, considering the dependencies between services.

\subsubsection{Temporal Features}\label{sec:temporal-features}
Temporal features are crucial in auto-scaling, as they enable proactive adjustments to resource allocation based on predicted future workloads, reducing performance degradation.

For workload prediction, we use a Long Short-Term Memory (LSTM) model, effective at capturing time-series data and long-term dependencies. The model is trained on the WorldCup98 dataset\cite{b16}, following the structure outlined in \cite{b20}, and achieves similar accuracy in forecasting workload fluctuations. By predicting traffic changes, the LSTM enables the system to adjust resources ahead of demand. LSTM's strength lies in its ability to handle sequential data and capture temporal dependencies, which traditional methods might miss.

Each STPID controller unit reads the predicted workload value during the auto-scaling decision process, using this information to adjust resource allocation accordingly. This ensures that the system can efficiently handle fluctuating traffic and reduce the risk of SLO violations.

\subsection{SpatioTemporal PID (STPID) Controller Units}
For each microservice in the application, we deploy a dedicated SpatioTemporal PID (STPID) controller unit. Each STPID unit consists of two key components: the \textbf{PID controller} and the \textbf{supervisory unit}. In the following subsections, we describe the individual roles of the PID controller and the supervisory unit in more detail.



\subsubsection{PID Controller}\label{sec:pid-controllers}
To dynamically adjust the resource allocation for each microservice in response to real-time performance data, we employ PID (Proportional-Integral-Derivative) controllers \cite{b17}. The PID controller is a widely used method in control systems due to its ability to balance quick response times, smooth adjustments, and long-term stability.

Each service \( S_i \) is equipped with a PID controller that adjusts the allocated resources based on the real-time measurement of the SLO violation \( \delta(S_i) \). The control action is calculated as follows\cite{b18,b19}:

\begin{equation}
\label{eq:pif1}
r_i(t+1) = r_i(t) + K_P \cdot e(t) + K_I \cdot \int_0^t e(t) \, dt + K_D \cdot \frac{de(t)}{dt}
\end{equation}

where:
\begin{itemize}
    \item \( r_i(t) \) is the control signal (the adjustment in resources) for service \( m_i \) at time \( t \),
    \item \( e(t) = \text{SLO}(S_i) - \delta(S_i) \) is the error (the deviation from the desired SLO) at time \( t \),
    \item \( K_P \), \( K_I \), and \( K_D \) are the proportional, integral, and derivative gains, respectively.
\end{itemize}

The PID controller continuously adjusts the resource allocation \( r_i(t) \) for each microservice to reduce the SLO violation \( \delta(S_i) \). In contrast to HPA, which cannot directly account for SLOs such as response time, PID controllers consider the entire service chain's SLO, enabling more precise and adaptive scaling decisions. As noted, each service has different specifications (e.g., resource needs, dependencies, and boot times), and these must be incorporated into the control mechanism. To address this, we extend the basic PID controller by integrating \textbf{spatiotemporal features} into the decision-making process. 




\subsubsection{Supervisory unit}\label{sec:spatial-features}
In a microservices architecture, each service has unique spatial features, such as its role, dependencies, and resource needs, which must be considered during resource allocation. To integrate these features into the auto-scaling process, we introduce a \textbf{supervisory mechanism} atop the PID controllers, ensuring that resource allocation accounts for three key factors:

\textbf{\textit{Service Weighting}}:
To efficiently allocate resources and minimize SLO violations, we assign a weight to each service based on its role in the service chain and its individual resource requirements. The weight of a service is determined by two main factors: its role in the service chain and its specific resource needs. Services that act as intersections or hubs between multiple service chains require more resources due to their increased load, and therefore are assigned higher weights. Additionally, services with higher resource demands—such as CPU, memory, and I/O operations—are also assigned higher weights. These resource requirements are derived from historical performance data, load testing, or application profiling.

To update the service weights, the supervisory unit adjusts them based on the service's recent performance and resource utilization, as provided by the global supervisory unit. If a service's CPU utilization has been below the LOW-THRESHOLD for the last 5 minutes (both of which are framework inputs), the weight assigned to the service is decreased, indicating performance issues with other services in the chain. Conversely, if the service's CPU utilization exceeds the HIGH-THRESHOLD (another framework input), the weight is increased, reflecting the need for more resources. These adaptive weight adjustments optimize resource allocation and ensure that the system responds effectively to dynamic workload conditions.

\textbf{\textit{Service Timing}}:
Each service has unique timing characteristics, including its startup time and resource provisioning requirements. The supervisory mechanism adjusts the PID controller's timing parameters to ensure smooth scaling and minimize delays during resource adjustments.

\textbf{\textit{Service Dependencies}}:
Services in a chain often depend on one another, and the supervisory mechanism accounts for these dependencies when adjusting resource allocation. It prioritizes resources for services that impact the performance of others, preventing bottlenecks. The mechanism adjusts resource allocation based on both individual service performance and the performance of related services. For example, if a service faces an SLO violation due to insufficient resources, the mechanism may increase resources for that service while avoiding unnecessary over-provisioning for neighboring services that are not experiencing issues.

\section{Evaluation}\label{sec:evaluation}
In this section, we present the evaluation of our proposed method for adaptive auto-scaling in microservices environments. We first describe the experimental setup, including the benchmark, workloads, and setup used for evaluation. Then, we present and analyze the results of the experiments to demonstrate the effectiveness of our framework in minimizing SLO violations and optimizing resource usage.

\begin{figure*}[t]
\centering
\includegraphics[width=0.9\textwidth]{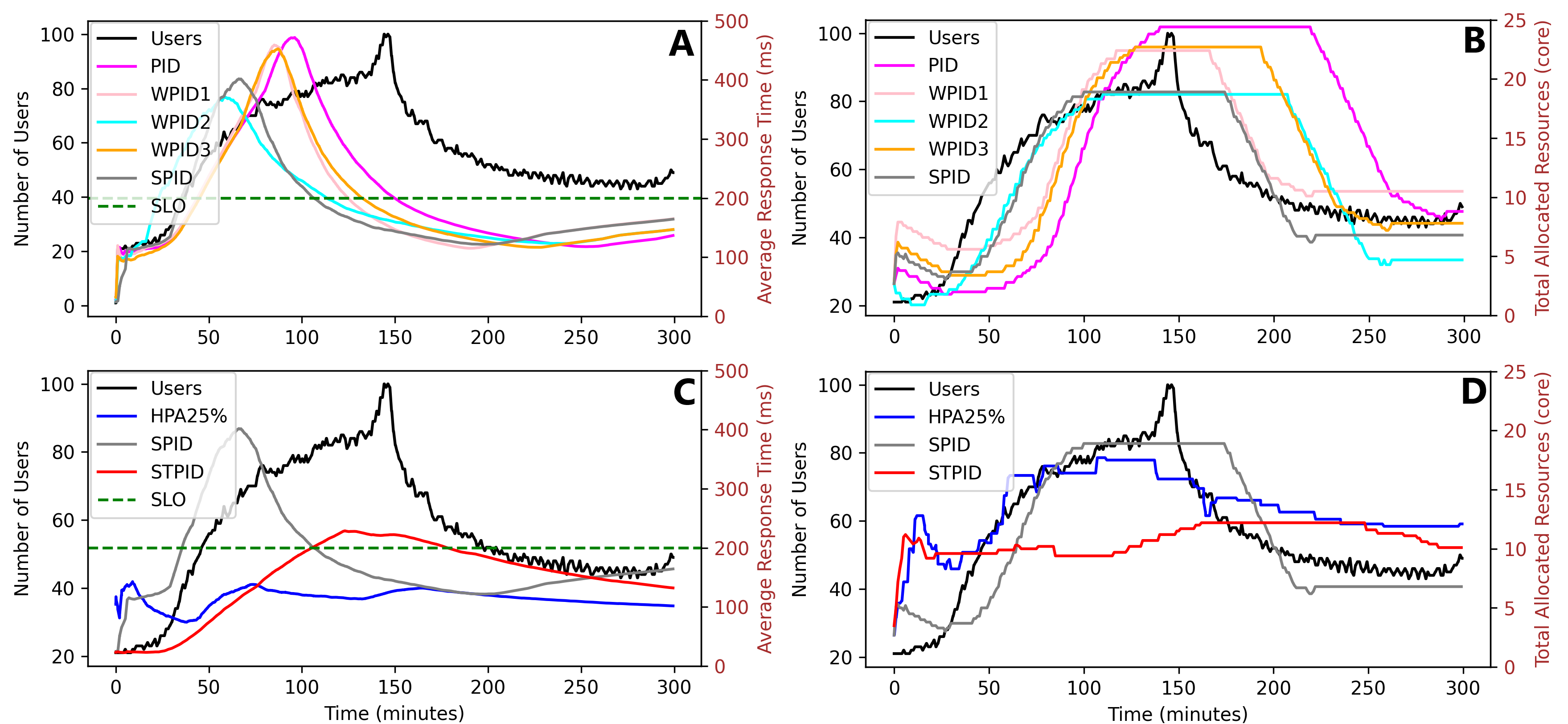}
\label{fig=STPIDRes}
\caption{A) RQ1: PID vs WPID vs SPID - Response Time B) RQ1: PID vs WPID vs SPID - Resource Usage, C) RQ2: HPA vs SPID vs STPID - Response Time D) RQ2: HPA vs SPID vs STPID - Resource Usage}
\end{figure*}

\subsection{Experiment Setup}\label{sec:experiment-setup}
\subsubsection{Benchmark}
For our experiments, we have deployed a demo microservice application, SockShop\cite{b12}, which is a well-known benchmark for microservices architectures. The application has been deployed on a self-hosted Kubernetes cluster, and we focus on evaluating the auto-scaling of the \texttt{/login} service, which includes services such as \texttt{front-end}, \texttt{user}, \texttt{user-db}, \texttt{carts}, and \texttt{carts-db}. 

\subsubsection{Workload}
For the experiment, we use the WorldCup98 dataset\cite{b16}, which records web traffic during the 1998 FIFA World Cup. This dataset captures both regular and peak load periods, providing a realistic simulation of traffic fluctuations for testing dynamic scaling.

To generate synthetic traffic, we utilize the LOCUST.io\cite{b15} load testing tool, which simulates real user behavior. Configured with the WorldCup98 dataset, LOCUST.io replicates real-world traffic patterns in the SockShop application, allowing us to assess the system's response to varying demand levels.

\subsubsection{Setup}
The experiments were conducted on a powerful server with dual AMD EPYC 7742 processors, 256 cores, and 1 TB of RAM. This server was divided into three Virtual Machines (VMs) to simulate the control plane and worker nodes in the Kubernetes cluster. The Kubernetes cluster, deployed on these VMs, provided a scalable environment for simulating real-world workloads. 

\subsection{Experiment Results and Analysis}

In this section, we evaluate the proposed method using the experiments described earlier, addressing the research questions (RQ1 and RQ2). The goal of this evaluation is to understand how the integration of spatial features and temporal features (such as workload prediction) contributes to the effectiveness of auto-scaling in microservices architectures.

\subsubsection{RQ1: Impact of Spatial Features on Auto-Scaling}

To answer RQ1, we evaluate the impact of incorporating spatial features—such as service roles within the service chain and service timing—on the auto-scaling performance of microservices. In the first step, we adjust the PID controller’s loop timing based on each service's boot time.

For the PID controller, we initially used the parameter values suggested in \cite{b19}, including \(K_P = 0.0004\), \(K_I = 0.0004\), and \(K_D = 0.00005\). While these values were a good starting point, they did not fully align with the needs of our microservice application. After experimentation and manual tuning, we obtained more appropriate values for the PID controller: \(K_P = 0.004\), \(K_I = 0.004\), and \(K_D = 0.0005\). For the first round of experiments, we used these tuned PID controller values for each service, including 'front-end', 'user', and 'carts'.

In the second round, we assigned a weight to each service and PID (WPID) controller based on resource requirements. Based on our preliminary experiments, we assigned resource weights ranging from 0.5 to 2, as we found that some services, such as \texttt{carts}, require up to twice the resources of others. In the final round, we applied our adaptive method to dynamically adjust the assigned weights for each PID controller based on real-time performance feedback.

Figure 3.A shows the response time for these three approaches, while Figure 3.B illustrates the resource usage for each method. Table II summarizes the results, comparing response time violations (in milliseconds) and total resource usage (in cores/min). The results indicate that assigning weights to the services improves SLO violations compared to the fixed PID controller approach, with a 15.6\% reduction in resource usage and a 59.0\% improvement in response time violations. Moreover, adaptively adjusting the weights further enhances SLO violations, leading to a greater improvement in both resource usage and response time violations, surpassing the fixed weight method.

The integration of spatial features led to better resource utilization, with fewer over-provisioning and under-provisioning instances. The PID controllers dynamically adjusted the resource allocation based on the spatial characteristics, ensuring that critical services were given priority in resource allocation.

\begin{tcolorbox}
Answer to RQ1: Incorporating spatial features, such as service roles and resource requirements, improved auto-scaling performance. Assigning and adaptively adjusting service weights reduced resource usage by 15.6\% and improved response time violations by 59.0\%.
\end{tcolorbox}

\subsubsection{RQ2: Impact of Spatial and Temporal Features on Auto-Scaling}

For RQ2, we explore the combined impact of spatial features and temporal features (workload prediction using LSTM) on the auto-scaling decision-making process. This part of the experiment examines whether incorporating predictions about future workloads further improves scaling decisions beyond just considering spatial dependencies. We refer to the PID controller with spatial features as SPID and the controller incorporating both spatial and temporal features as STPID. 

Figure 3.C shows the response time for both SPID and STPID, while Figure 3.D illustrates the resource usage for both. Table 2, rows 5 and 6, summarizes the response time and resource usage for SPID and STPID. The results show a significant improvement in both SLO violations and resource usage with the integration of temporal features, in addition to spatial features, with STPID outperforming SPID by 11.70\% in resource usage and 76.2\% in response time violations.


\begin{tcolorbox}
Answer to RQ2: Integrating temporal features (workload prediction) with spatial features significantly improves auto-scaling performance. The STPID controller outperforms SPID, reducing resource usage by 11.7\% and improving response time violations by 76.2\%.
\end{tcolorbox}

\begin{table}[htbp]
\caption{Comparison of Resource Usage and Response Time Violations for Uniform and Differentiated Scaling Approaches}
\begin{center}
\begin{tabular}{|c|c|c|c|c|c|}
\hline
\textbf{}&\multicolumn{3}{|c|}{\textbf{PID weights }}& \textbf{Resource} & \textbf{Violations} \\
\cline{2-4} 
\textbf{} & \textbf{Front}& \textbf{User}& \textbf{Carts} & (Core/Min) & (ms)\\
\hline

HPA25 & - & - & - & 4039 & 0 \\
\hline
PID & 1 & 1 & 1 &  3963 & 64651  \\
\hline
WPID1 & 0.5 & 2 & 1  & 3897 & 48161 \\
\hline
WPID2 & 1 & 0.5 & 2  & 3509 & 38330 \\
\hline
WPID3 & 2 & 1 & 0.5  & 3866 & 50355 \\
\hline
SPID & * & * & * &  3346 & 26521 \\
\hline
STPID & * & * & * & 2954 & 6309 \\
\hline

\multicolumn{4}{l}{$^{*}$Adaptive}

\end{tabular}
\label{HPA_Results}
\end{center}
\end{table}

\section{Related Work} Auto-scaling of microservices has gained significant attention due to the need for dynamic resource allocation in cloud environments. This section reviews research in three key areas: spatial features, temporal features, and spatiotemporal approaches in microservice auto-scaling.

\subsection{Spatial Features in Auto-scaling:} Xie et al. \cite{b7} propose PBScaler, a bottleneck-aware framework that optimizes scaling decisions by identifying performance bottlenecks, improving resource consumption. Meng et al. \cite{b23} present DeepScaler, a deep learning-based approach that addresses service dependencies, reducing SLA violations by 41\%. Yu et al. \cite{b24} introduce Microscaler, which uses service dependency graphs and a hybrid learning approach to optimize resource allocation, achieving high precision and faster convergence. Baresi et al. \cite{b25} introduces NEPTUNE\textsuperscript{+}, a resource allocation solution for serverless functions on edge infrastructures that considers function dependencies. The solution is based on a theoretical model and a control algorithm designed to account for these dependencies

\subsection{Temporal Features in Auto-scaling:} Temporal features focus on predicting and adapting to workload fluctuations. Straesser et al. \cite{b21} introduce a decentralized autoscaling approach that improves adaptability in dynamic environments. Bai et al. \cite{b22} propose a distributed autoscaling solution using Twin Delayed Deep Deterministic Policy Gradient, which reduces response times in large-scale clusters. Quattrocchi et al. \cite{b8} compare various autoscaling solutions, showing that their proposed methods outperform cloud providers in SLA violations and resource usage. Nunes et al. \cite{b6} propose MS-RA, a self-adaptive autoscaling solution that outperforms Kubernetes HPA by meeting SLOs with fewer resources. Sabuhi et al. \cite{b19} combine neural networks with a PID controller for adaptive autoscaling, ensuring SLA compliance and efficient resource use.

\subsection{Spatiotemporal Approaches in Auto-scaling:} Li et al. \cite{b4} introduce DCScaler, a collaborative autoscaler using spatiotemporal predictions and deep reinforcement learning to improve resource utilization and reduce SLA violations. Liao et al. \cite{b5} present STAAF, a proactive autoscaling framework that reduces tail latency by 74\% during traffic surges by considering spatial-temporal correlations. Li et al. \cite{b26} introduce MarVeLScaler, which dynamically scales cloud clusters for MapReduce jobs, reducing costs by 30.8\% while maintaining performance.

\section{Conclusion and Future Work}\label{sec:conclusion}

In this paper, we proposed a framework to improve microservice auto-scaling by considering spatial and temporal features. To the best of the author's knowledge, this is the first work to use three orthogonal research dimensions—control theory, machine learning, and intelligent heuristics—to address spatiotemporal-aware adaptive auto-scaling for cloud-based microservice applications. We demonstrated that incorporating spatial features, such as service dependencies and resource requirements, helps optimize resource allocation. Furthermore, integrating temporal features using LSTM for workload prediction enhanced performance by anticipating future workload fluctuations, thus reducing SLO violations. Our experimental results show that while the STPID approach did not reduce response time violations compared to HPA (violations for STPID are 6309 ms, while HPA has 0 ms), the resource usage was decreased by 27\%, with HPA consuming 4039 cores/min and STPID consuming 2954 cores/min. The violation level of STPID is close to that of HPA with a 25\% improvement, demonstrating that the integration of temporal features helps optimize resource usage without significantly worsening performance.

For future work, several areas can be explored to further enhance our framework. In terms of spatial features, we plan to use machine learning models to predict future response times and adaptively adjust PID controller weights. Additionally, we aim to test multiple service chains concurrently. For temporal features, we intend to explore multi-step LSTM models to improve the accuracy of workload predictions.

From an experimental perspective, we plan to test our solution in a larger cluster with more nodes, use public cloud providers, and incorporate more diverse traffic patterns, including sharp spikes. We will also investigate the impact of sending concurrent traffic to multiple services.

In conclusion, this work provides a foundation for improving auto-scaling in microservice architectures and sets the stage for future research aimed at further optimizing scalability and performance in dynamic cloud environments.

\end{document}